\documentclass[runningheads]{llncs}
\usepackage{graphicx}
\usepackage{wrapfig}
\usepackage{algorithm}
\usepackage{algorithmic}
\usepackage{prettyref}
\usepackage{listings}
\newrefformat{lst}{Listing \ref{#1}}
\newrefformat{alg}{Algorithm \ref{#1}}

\lstdefinelanguage{json}{
    basicstyle=\small\ttfamily,
    numbers=none,
    numberstyle=\scriptsize,
    stepnumber=1,
    numbersep=8pt,
    showstringspaces=false,
    breaklines=true,
    frame=lines,
    literate=
     *{0}{{{0}}}{1}
      {1}{{{1}}}{1}
      {2}{{{2}}}{1}
      {3}{{{3}}}{1}
      {4}{{{4}}}{1}
      {5}{{{5}}}{1}
      {6}{{{6}}}{1}
      {7}{{{7}}}{1}
      {8}{{{8}}}{1}
      {9}{{{9}}}{1}
      {:}{{{{:}}}}{1}
      {,}{{{{,}}}}{1}
      {\{}{{{{\{}}}}{1}
      {\}}{{{{\}}}}}{1}
      {[}{{{{[}}}}{1}
      {]}{{{{]}}}}{1},
}

\begin{document}
%
\title{Enabling EASEY deployment of containerized applications for future HPC systems}
\titlerunning{EASEY containerized applications}
%
\author{Maximilian H\"ob
\and Dieter Kranzlm\"uller
}
\authorrunning{M. H\"ob, D. Kranzlm\"uller}
%
\institute{MNM-Team, Ludwig-Maximilians-Universit\"at M\"unchen \\ Oettingenstraße 67, 80538 M\"unchen, Germany \\
\email{hoeb@mnm-team.org}
\url{www.mnm-team.org}}
\maketitle              
\begin{abstract}

The upcoming exascale era will push the changes in computing architecture from classical CPU-based systems towards hybrid GPU-heavy systems with much higher levels of complexity. While such clusters are expected to improve the performance of certain optimized HPC applications, it will also increase the difficulties for those users who have yet to adapt their codes or are starting from scratch with new programming paradigms. Since there are still no comprehensive automatic assistance mechanisms to enhance application performance on such systems, we propose a support framework for future HPC architectures, called EASEY (Enable exASclae for EverYone). Our solution builds on a layered software architecture, which offers different mechanisms on each layer for different tasks of tuning, including a workflow management system. This enables users to adjust the parameters on each of the layers, thereby enhancing specific characteristics of their codes. We introduce the framework with a Charliecloud-based solution, showcasing the LULESH benchmark on the upper layers of our framework. Our approach can automatically deploy optimized container computations with negligible overhead and at the same time reduce the time a scientist needs to spent on manual job submission configurations.

\end{abstract}

\section{Introduction}

Observation, Simulation and Verification build the pillars of most of today's HPC applications serving different goals of diverse scientific domains. Those applications have changed in the last decades and years, and they will and have to change again, driven by several factors. More applications, more scientists, more levels of detail, more data, more computing power, more of any contributing part. This more of everything needs to be satisfied by current and future computing systems, including not only computing power, storage and connectivity, but also direct application support from computing centers.

Such a support is essential to execute a demanding high performance application with huge data sets in an efficient manner, where efficiency can have several objectives like time to solution or energy consumption. The latter will be the crucial factor to minimize to achieve an exascale system that serves the scientific community and does not stress our environment.

Computing efficiency in means of likely optimal usage of resources in an acceptable time to solution is heavily investigated from different sites. Many workflow management frameworks promise enhancements on data movement, deployment, management or reliability, like in SAGA, a grid based tool suit described in \cite{goodale2006saga} or Pegasus, a scientific pipeline management system presented in \cite{deelman2015pegasus}. The scalability of such frameworks is limited by the state-of-the-art services, where bottlenecks are found inside and across today's supercomputing facilities. 

Solutions to this challenge will not be found in one single place. Instead, it will be the most optimal interaction of exascale-ready services, hardware and applications. To support todays and future application developers, automatic assistant systems will be a key enabling mechanism, also shown by Benkner et al. in \cite{benkner2014automatic}. Computing systems will continue to develop and change faster than applications can be adapted to build likely optimal synergies between the application and the underlying system. Heterogeneity among already connected centers will additionally increase the complexity.

Although scientist want to simulate, calculate, calibrate or compare data or observations always with the same application core, many hurdles slow down or stop those scientist to deploy on newer systems, since the effort to bring their application on new hardware or on new software stacks is too high in comparison to their actual work in their scientific domains. An astro physicist needs to focus on physical problems, not on how to deploy applications on a computing cluster. 

This is supported by a survey from Geist and Reed in \cite{geist2017survey} who state, that the complexity of software needs to be reduced to reduce software development costs. This could also be solved if we can encourage application developers to include building blocks, which will adapt and optimize code (compare \prettyref{sec:future}) or executions automatically, like proposed in our paper. 

Therefore, new assistent systems are needed to make access to new supercomputing centers easier and possible also for unexperienced scientist. The complexity of those systems requires knowledge and support, which usually only the computing centers themselves can offer. Since their time is limited also the diversity of the applications might be limited.

This work introduces a framework to enable container applications based on Docker to be transformed automatically to Charliecloud containers and executed on leading HPC systems by an integrated workflow management system. This transformation includes also the possibility to define mounting points for data. Charliecloud is considered the most secure container technology for supercomputers, since Singularity is not entirely free of breaches as reported in \cite{CVE-2019-11328}. 

In this work we propose initial steps towards the first comprehensive framework, already including auto-tuning mechanisms focused on containerized applications. Using containers to encapsulate an application with all its dependencies and libraries introduces portability in general. With EASEY also specific libraries can be added to the portable container to optimize the performance on a target system automatically.

Within this paper we introduce the underlying technology and present the functionality of the framework, evaluated by a hydrodynamics stencil calculation benchmark. We show, that our approach adds automatically cluster dependent building bricks, which improve the utilization of the underlying hardware only by acting on the software layer of the container. With this auto-tuning mechanism, we reduce the necessary time domain scientists need to invest in deploying and managing their HPC jobs on different clusters and can in stead concentrate on their actual work in their domain.

The paper is ordered as follows. \prettyref{sec:easey-arch} introduces the architecture of our framework integrated into the layered architecture of supercomputing systems. Afterwards, \prettyref{sec:easey-conf} presents the necessary configuration needed for EASEY. \prettyref{sec:eval} evaluates this approach with a benchmark use case. Related Work to this paper is presented in \prettyref{sec:related} and \prettyref{sec:conclusion} closes with a summary and an outlook of future work to extend this approach to other layers of the HPC architecture.

\section{EASEY Architecture}
\label{sec:easey-arch}

Enabling scientists to focus on their actual work in their domain and remove all deployment overhead from their shoulders is the main goal of tour approach. And while we reduce the necessary interaction between scientist and compute cluster, we also apply performance optimization on the fly. High performance systems need applications to adapt their technology to talk their language. We are able to add such optimizations while preparing the containerized application.

\begin{figure}
\center
\includegraphics[width=1\linewidth]{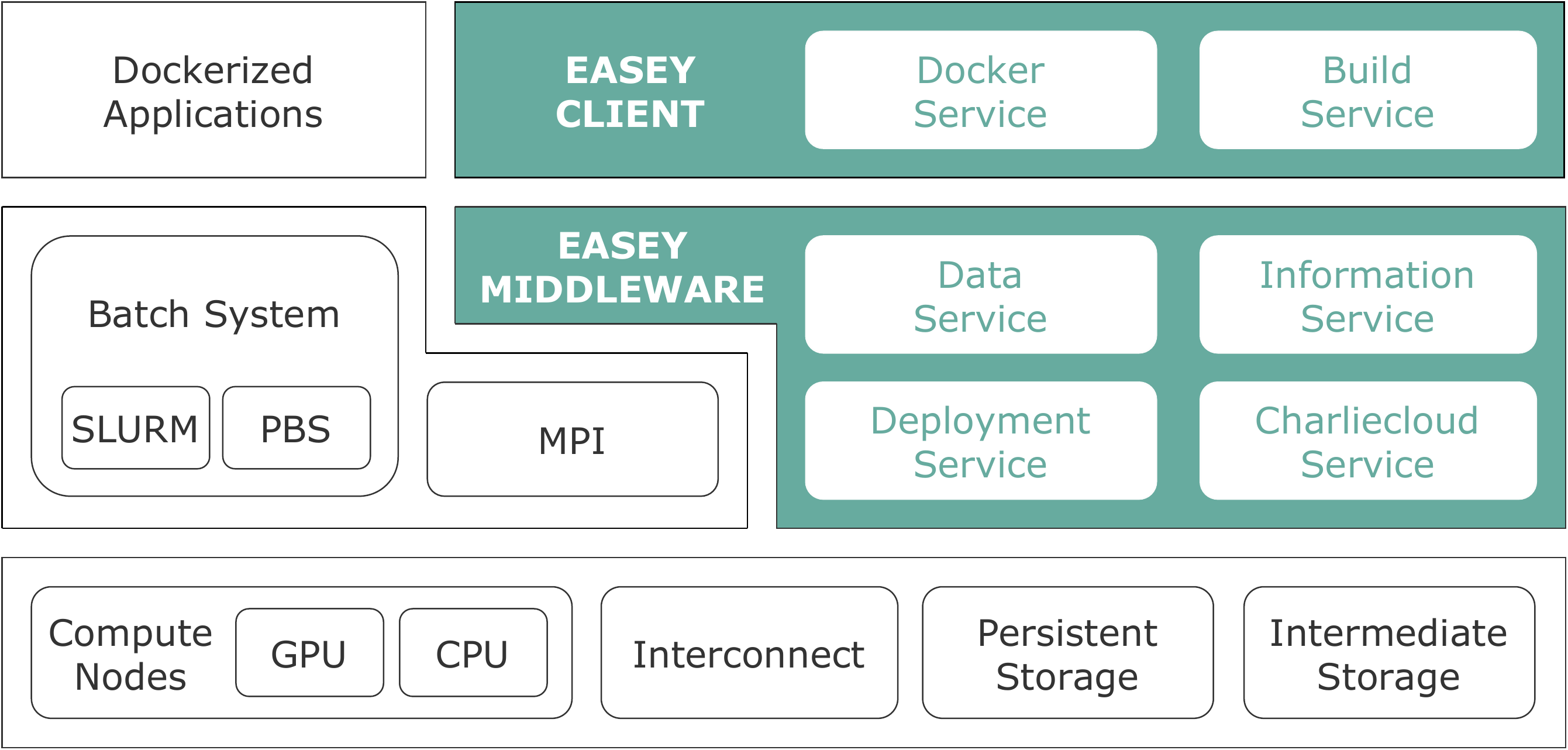}
\caption{EASEY integration in the layered architecture of HPC systems} \label{fig:easey-stack}
\end{figure}

The architecture of the EASEY system is detailed in \prettyref{fig:easey-stack}, integrated as two building bricks in the layered HPC architecture. On the upper \textit{Applications and Users layer}  the EASEY-client is mainly responsible for a functional build based on a Dockerfile and all information given by the user. The middleware on the \textit{local resource management layer} takes care of the execution environment preparation, the data placement and the deployment to the local scheduler. The additional information service can be pulled for monitoring and status control of the execution through. The \textit{hardware layer} of the compute cluster underneath remains not included in any optimization in this release of the framework (compare future work in \prettyref{sec:conclusion}).

\subsection{EASEY Client}

The client  as the main service for any end-user prepares the basis of the execution environment by collecting all needed information to build a Charliecloud container. Therefore, the main information is given by the user with the Dockerfile. This file can also be pulled from an external source and needs to include all necessary steps to create the environment for the distinguished tasks. The client needs at some point root privileges to build the final container, hence, it needs to be deployed on a user's system like a workstation or a virtual machine in a cloud environment.

As a first step, the \textit{Docker Service} builds a docker image. Since HPC clusters can require local configurations, the \textit{Docker Service} adds local dependencies to the Dockerfile. Therefore, the user needs to specify the target system in the build command:  \lstinline[basicstyle=\small\ttfamily]{easey build Dockerfile --target cluster}, e.g. \textit{easey build Dockerfile {-}{-}target "lrz:supermuc-ng"}.

In addition for mpi-based applications, the actual mpi-version needs to match the system's version. In the Dockerfile the user can specify the position where the cluster's mpi-version needs to be integrated by including the line \linebreak \lstinline[basicstyle=\small\ttfamily]{###includelocalmpi###}, which will be replaced by the client with the actual purge of all other mpi-versions and the compilation of the needed one. This should be done before the target application is compiled to include the right mpi libraries.
 
 As a final step the later mounting point will be created as a folder inside the Docker image. The path was defined inside the configuration file (see \prettyref{lst:data}). Also specific requirements from the target system will be handled here, for example to include local libraries and functionalities inside the container (e.g. symlinks to special folders). Those requirements are known by the EASEY system and don't need to be provided by the user.

In the same environment as the \textit{Docker Service} the \textit{Build Service} will transform the before created Docker image to a Charliecloud container archive. The service will call the Charliecloud command \lstinline[basicstyle=\small\ttfamily]{ch-builder2tar} and specify the Docker image and the build location.

\subsection{EASEY Middleware}

The second building brick is the EASEY Middleware, which connects and acts with the resource manager and scheduler. The main tasks are \textit{job deployment},  \textit{job management}, \textit{data staging} and creating the \textit{Charliecloud environment}.

\begin{figure}
\center
\includegraphics[width=1\linewidth]{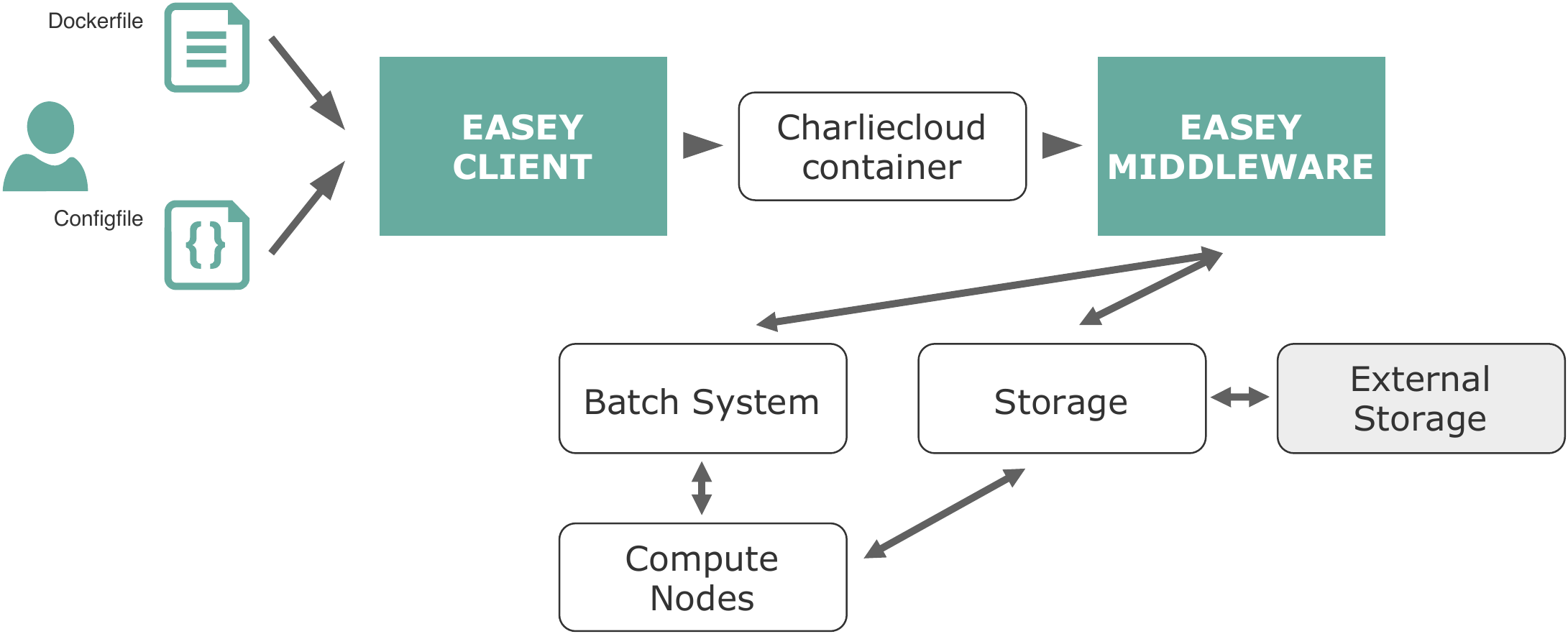}
\caption{EASEY workflow of the job submission on HPC systems} \label{fig:easey-workflow}
\end{figure}

Thereby, a workflow is started including besides the before created Charliecloud container and configuration file, the local cluster storage and batch system as well as the user specified external storage, if needed. A schematic view of the major steps is given in \prettyref{fig:easey-workflow}. 

Starting with the user, the Dockerfile and the filled EASEY configuration file need to be included in a build call of the EASEY client, which is running on a user system with root privileges. Within this process a Docker container is built and transformed to a Charliecloud container, which again is packed in a tar-ball. 

The EASEY middleware can be placed inside the target cluster or outside on a virtual machine for example. The framework will start the preparation of the submission based on the information given in the configuration. This can also include a data stage-in from external sources. To place data and later to submit a cluster job on behalf of the user, EASEY needs an authentication or delegation possibility on each contributing component. At this time of the development the only possibility included is access grants via public keys. This means in detail, if the EASEY middleware runs outside the cluster, that the public key of the host system needs to be added to the \textit{authorized keys} inside the cluster. Thereby, EASEY can transfer data on the user's storage inside the cluster. 

Also the following job deployment needs a working communication from the middleware to the cluster submission node. The deployment based on the given configuration (see \prettyref{lst:execution}) follows a well reinforced approach. The complete algorithm is shown in \prettyref{alg:submission}.

\begin{algorithm} 
\caption{EASEY submission} 
\label{alg:submission} 
\begin{algorithmic}
    \REQUIRE Charliecloud tar-ball
    \REQUIRE EASEY configuration file
    \REQUIRE User credentials
    \STATE Move tar-ball to cluster storage
    \STATE Extract tar-ball and create execution environment
    \IF{$data$ in configuration}
      \STATE mkdir data\_folder
    \ENDIF
    \WHILE{$input$ in configuration}
        \STATE transfer $input[source]$ to data\_folder
    \ENDWHILE
    \STATE create batch\_file
    \FOR{each $deployment$ in configuration}
        \STATE parse to SLURM or PBS command in batch\_file
    \ENDFOR
    \WHILE{$execution$ in configuration}
        \STATE add command to batch\_file
    \ENDWHILE
    \STATE submit batch\_file to local scheduler and return $jobID$ to EASEY
\end{algorithmic}
\end{algorithm}

Additionally to the already mentioned tasks the \textit{data folder} is only created if at least one input or output file in specified. EASEY requires the user to place the output file after the computations inside this folder, mounted inside the Charliecloud container.  For each input file, EASEY controls the transfer inside the data folder. 

For the submission on the local batch system a batch file is required following also local specifications, known by EASEY. The resource allocation information is provided by the user in the configuration (\textit{number of nodes}, \textit{ram}, ...). For SLURM or PBS those are parsed into a valid form, other scheduler are not supported so far. 

The actual computations follow after this prolog, described as \textit{executions} by the user. For each the corresponding bash or mpi commands are also included. If data is required as input parameters, the user has to specify them relatively to the data folder, where they are placed.

Since the local job ID is known by the middleware the user can pull for the status of the job. In addition to \textit{pending}, \textit{running}, \textit{finished} or \textit{failed}, also error log and standard output is accessible, also at an intermediate state. After the job ended EASEY will transfer output files if specified. 

This workflow includes all necessary steps to configure and run an originally Docker based application on a HPC cluster. Thereby, it saves time any scientist can use for actual work in their domain and removes any human overhead especially if such computations need to be deployed regularly. In the same time, it adds optimization mechanisms for the actual computing resource. In the following section, details on the user's mandatory configuration are presented.

\section{EASEY Configuration}
\label{sec:easey-conf}

Our approach requires a full and valid description of all essential and optional parts of the submission. Therefore we defined a json-based configuration file including all required information. This file needs to be provided by the user together with the application's Dockerfile. The configuration consists of four main parts: \textit{job}, \textit{data}, \textit{deployment} and \textit{execution}.

\paragraph{Job Specification}

This part of the configuration can also be seen as mandatory meta data for a valid job management. The keys of the key-value pairs are presented in \prettyref{lst:job}.

An EASEY job needs to have an unique identifier, a hash which is determined by the system at the moment of submission, a user specified \textit{name} and a \textit{mail address} to contact the end-user if specified. Besides those, no further information is mandatory.

  \hspace*{-\parindent}%
\begin{minipage}{0.48\linewidth}
\begin{lstlisting}[language=json,firstnumber=1,caption={Job Specification}, label={lst:job}]
{"job":{"name","id","mail",
 "data":{..}, 
 "deployment":{..},
 "execution":{..}}
}
  \end{lstlisting}
    \vspace{0.74cm}
  \end{minipage}
  \hspace{0.01\linewidth}
\begin{minipage}{0.48\linewidth}
\begin{lstlisting}[language=json,firstnumber=1,caption={Data Specification}, label={lst:data}]
"data":{"input":[
  {"source","protocol", 
  "user","auth"}],
"output":[
  {"destination","protocol", 
  "user","auth"}],
"mount":{"container-path"}}
\end{lstlisting}

  \end{minipage}

\paragraph{Data Service Specification}

Our backend is able to fetch accessible data files via the protocols https, scp, ftp and gridftp. The latter is planed to be implemented in the next release. For the others already available only the path to the source and the protocol needs to be declared. If the data needs to be accessed on a different site, authentication with public-key mechanism is necessary. The  \textit{input} is declared as an array and can include several declarations of different input files.

The backend will place all input files in one folder, which will be mounted into the container on the relativ position declared as \textit{path}.

After the complete execution an \textit{output} recommend as an archive can be also moved again to a defined destination. Also here a public key mechanism would be mandatory.
  
  \paragraph{Deployment Service Specification}

The deployment service offers basic description possibilities to describe necessary resources for the execution. 

As shown in the next section, within one deployment only one job is allocated. Therefore, each execution commands specified in \prettyref{lst:execution} will be run on the same allocation. The specifications regarding \textit{nodes}, \textit{ram}, \textit{taks-per-node} and \textit{clocktime} will be translated into scheduler specific commands and need to be specified given in \prettyref{lst:deployment}. 
   
  \hspace*{-\parindent}%
\begin{minipage}{0.48\linewidth}
\begin{lstlisting}[language=json,firstnumber=1,caption={Deployment Specification}, label={lst:deployment}]
"deployment":{"nodes",
 "ram","cores-per-task",
 "tasks-per-node","clocktime"
}
  \end{lstlisting}
  \vspace{0.8cm}
  \end{minipage}
  \hspace{0.01\linewidth}
\begin{minipage}{0.48\linewidth}
\begin{lstlisting}[language=json,firstnumber=1,caption={Execution Specification}, label={lst:execution}]
"execution":[{
 "serial":
  {"command"},
 "mpi":
  {"command","mpi-tasks"}
}]
\end{lstlisting}
  \end{minipage} 
  
  Although there exist much more possible parameters, at this state of the framework only those are implemented, since all others are optional.
   
       \paragraph{Execution Service Specification}   
    The main ingredients of HPC jobs are the actual commands. The \textit{execution} consists of in principle unlimited \textit{serial} or \textit{mpi} commands. Those are executed in order of sequence given inside the \textit{execution} array as shown in \prettyref{lst:execution}.. In all considered HPC jobs the only kinds of commands are bash (\textit{serial}) or \textit{mpi}-based commands.

A complete example is given in \prettyref{lst:lulesh} showing a practical description of the evaluated use case. The presented configuration will of course be adapted whenever necessary. However, the main goal is to stay as generic as possible to connect and enable as many combinations of applications on the one side and resource management  systems on the other. The next section evaluates this approach regarding the computational and the human overhead. 

\section{Evaluation}
\label{sec:eval}

The previously presented framework builds the basis for further development. The main goal is to introduce auto-tuning on several layers. In this paper, we presented the EASEY client and middleware to support scientists deploying a Docker-based application on a HPC cluster without interacting with the local resource themselves. 

This framework was tested on one of the fastest HPC systems in the world, the SuperMUC-NG, a general purpose system at the Leibniz Supercomputing Center\footnote{\url{https://www.lrz.de/english/}} in Garching, listed ninth in the Top500 list in November 2019 and has a peak performance of 26.87 Petaflops, computing on 305,856 Intel Xeon Platinum 8174 CPU cores, without any accelerators. All compute nodes of an island are connected with a fully non-blocking Intel Omnipath OPA network offering 100 Gbit/s, detailed in \cite{brayford2019deploying}. SuperMUC-NG uses SLURM as a system scheduler.

We used a Dockerimage for LULESH, the Livermore Unstructured Lagrangian Explicit Shock Hydrodynamics benchmark. It is a widely used proxy application to calculate the Sedov blast problem that highlights the performance characteristics of unstructured mesh applications. Details on the application and the physics are described by Karlin et al. in \cite{karlin2013lulesh}. 

This benchmark in version 2.0.3 was ported by the MNM research team (Fürlinger et al., described in \cite{Fuerlinger:2016:DASH}) to DASH, a C++ template library for distributed data structures, supporting hierarchical locality for HPC and data-driven science. Adopting the Partitioned Global Address Space (PGAS) programming model, DASH developed a template library that provides  PGAS-like abstraction for important data containers and allows a developer to control and take advantage of the hierarchical data layout of global data structures. The authors showed, that DASH offers a performance advantages of up to 9\%.

As described in \prettyref{sec:easey-conf} the EASEY client requires a Dockerfile and a configuration specifying the deployment, data and execution parameters. Since LULESH does not require any data, the json configuration shown in \prettyref{lst:lulesh} contains only \textit{job meta data}, \textit{deployment} and \textit{execution}. Values, which are determined by EASEY, or which are not defined by the user (e.g. \textit{ram} since there are no special memory requirements) are not set.

The actual execution is given by the \textit{command} keyword. In this case a charliecloud container is started with \textit{ch-run} and a data volume is mounted with the -b flag, \textit{-b source:target}. Inside the container \textit{lulesh.dash} the command \textit{/built/lulesh.dash -i 1000 -s 13} is executed. Together with the \textit{mpi-tasks} LULESH is ran with a cube size of 2.197 cells, a cube mesh length of 13, and in 1.000 iterations. The maximum runtime is limited to \textit{6 hours} and passed to the SLURM scheduler.

\begin{lstlisting}[language=json,firstnumber=1,caption={LULESH:DASH Execution Specification}, label={lst:lulesh}]
{"job":{
 "name":"LULESH:DASH","id":"",
 "mail":"hoeb@mnm-team.org", 
 "deployment":{
  "nodes":"46","ram":"","cores-per-task":"1",
  "tasks-per-node":"48","clocktime":"06:00:00"
 },
 "execution":{
  "serial": 
   {"command":"echo \"Starting LULESH:DASH\""},
  "mpi": 
   {"command":"ch-run -b /lrz/sys/.:/lrz/sys -w lulesh.dash -- /built/lulesh -i 1000 -s 13", 
    "mpi-tasks":"2197"},
  "serial": 
   {"command":"echo \"Finished LULESH:DASH\""},
 }
}
\end{lstlisting}

\begin{figure}
\centering
\begin{minipage}{0.48\linewidth}
\includegraphics[width=1\textwidth]{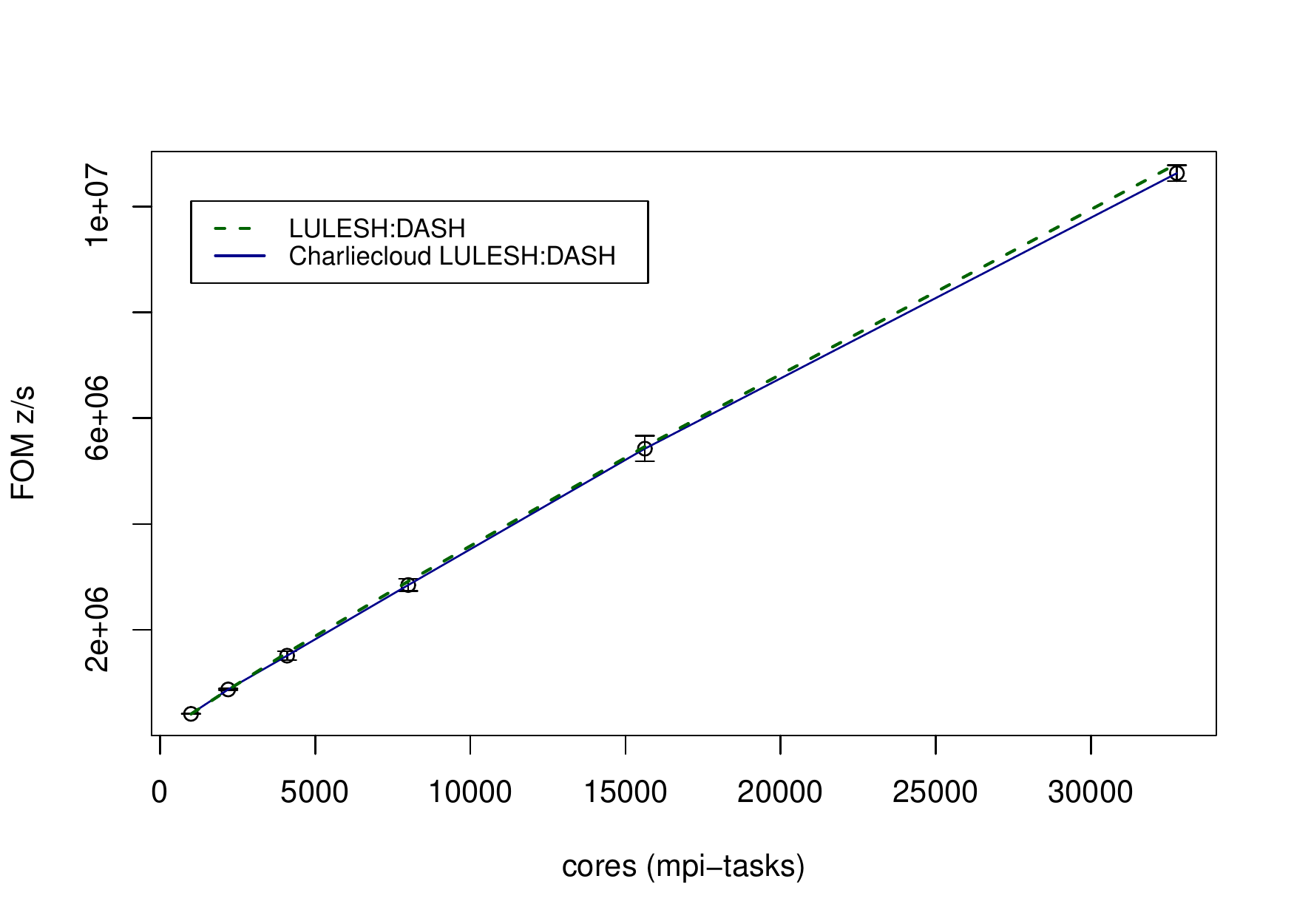}
\caption{Weak scaling on SuperMUC-NG} \label{fig:luleshscaling}
  \end{minipage}
\begin{minipage}{0.48\linewidth}
\includegraphics[width=1\textwidth]{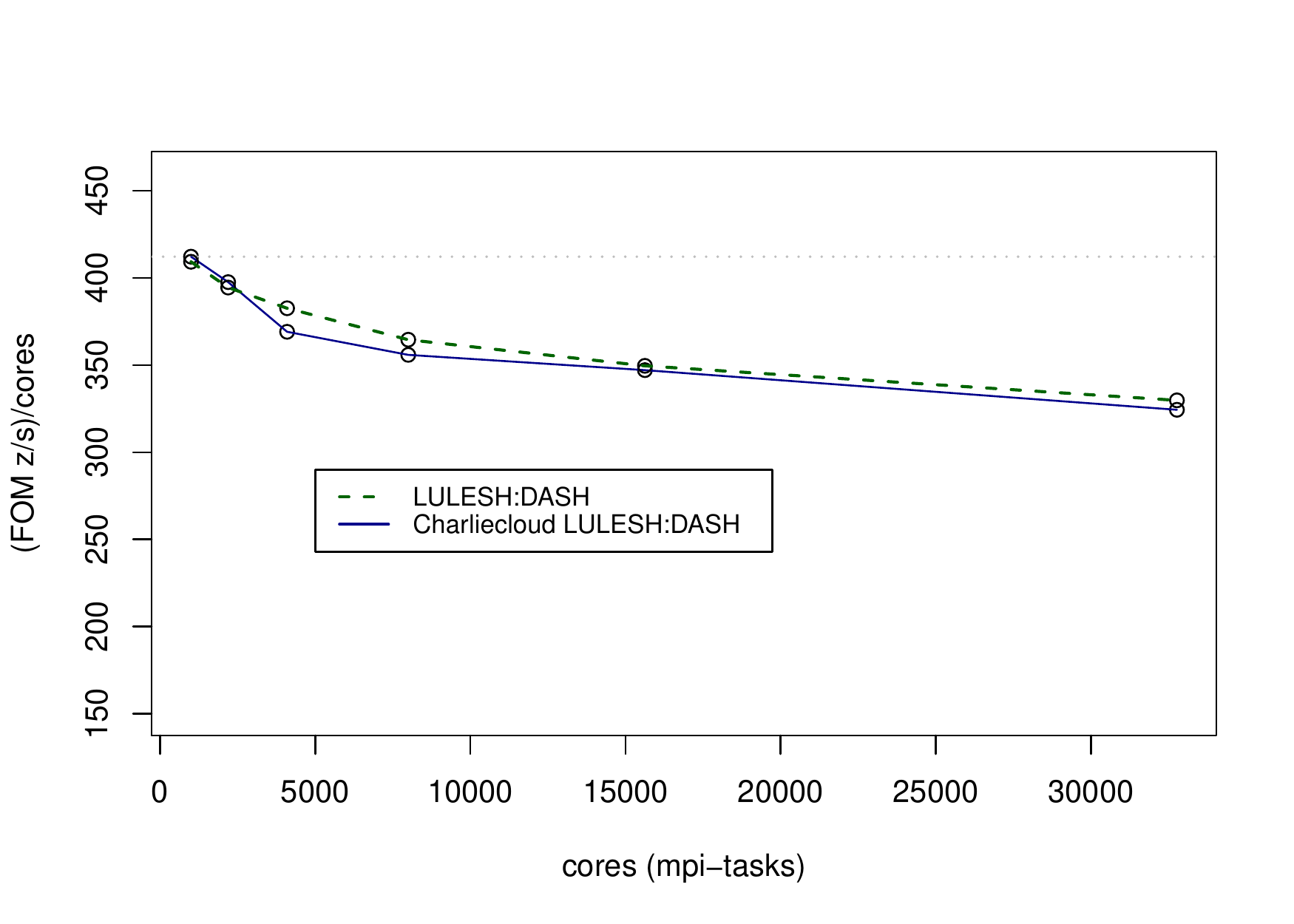}
\caption{FOM per cores SuperMUC-NG} \label{fig:luleshscaling2}

  \end{minipage} 
  \end{figure}

This setup was used to run several execution of the DASH LULESH and the DASH LULESH inside the Charliecloud container, on up to 32,768 cores. As it can be seen in \prettyref{fig:luleshscaling}, the figure of merit (FOM) shows slightly higher values (higher is better) for native DASH than for the Charliecloud runs. The FOM values of the Charliecloud executions are lower for runs with more than 4,000 cores. With less cores they differ under 1\% (compare \prettyref{tab:comparison}). This can be seen in detail in \prettyref{fig:luleshscaling2}, where the FOM value is divided through the number of cores. Ideally we would see a horizontal line, however, the difference between this line and the measurements corresponds to the application, which does not have perfect linear scaling. However, the scaling behavior of the containerized application is similar to the native one although some overhead introduced by Charliecloud is visible. 

The detailed mean measurements between the dash version of native LULESH and the EASEY container execution inside the Charliecloud container can be seen in \prettyref{tab:comparison}, where the number of cores (and mpi-tasks) correspond to cubic numbers of the given input cube side length. The shown delta varies from $+0,8 \%$ to $-3,6\%$ of the FOM values. This spread can be explained by the limited number of runs, that could be performed on the target system, the SuperMUC-NG of the LRZ, and statistical deviations. 

However, the usage of a Charliecloud container adds some overhead to the execution shown in the measurements with more the 4,000 cores. This overhead needs to be compared to the invested effort on executing this application on the system. With EASEY it was possible to execute and measure the application without manual interaction on the system itself. The so added performance overhead is within an acceptable interval. Especially for runs with many CPUs (10,000+) this overhead does not increase significantly. This is especially important, since our framework targets later Exascale systems and can already show today its scalability.

\begin{table}[]
\centering
\setlength{\tabcolsep}{7pt} 
\renewcommand{\arraystretch}{1.2}
\caption{FOM comparison: lulesh:dash native and inside charliecloud container.} 
\label{tab:comparison}
\begin{tabular}{|r|r|r|r|r|r|}
 \hline cube length $p$ & cores $p^3$ & nodes & FOM      EASEY & FOM  NATIVE& $\Delta$    \\ \hline
10&1,000  &           21&     412,122.1 &  409,204,8&	0,71 $\%$  \\ \hline
13&2,197  &         46&    873,366.4    &  866,515,2	&0,78 $\%$    \\ \hline
16&4,096  &          86&         1,511,665.1  & 1,566,899,9&	-3,65 $\%$ \\ \hline
20&8,000  &          167&     2,846,589.0    &  2,916,102,0	&-2,44 $\%$   \\ \hline
25&15,625  &         326&         5,423,072.1  &  5,461,509,5	&-0,71 $\%$ \\ \hline
32&32,768  &         683&          10,627,767.7 & 10,805,287,0	&-1,67 $\%$  \\ \hline
\end{tabular}
\end{table}

The goal of these measurements was not to show an optimal scaling behavior of the application, it was to demonstrate the validity of the approach. Although there might be some additional overhead due to Charliecloud, EASEY could reproduce the scaling behavior and very closely the overall performance of the original, manually compiled application without any container framework. This shows that the approach of EASEY adds only negligible overhead to the performance. In the same time it saves the scientist time by automatically tuning some adjusting screws.

With the weak scaling shown in \prettyref{fig:luleshscaling} we can show, that our approach scales as well as the manually compiled application without any container environment. Automatically enabling a container deployment on such a supercomputing cluster and in the same time applying local tuning possibilities show, that EASEY is a promising approach. It is also likely that such assistance systems will increase to number of users using those HPC systems and in the same time enabling them to include as much optimization as possible, without changing anything manually. 

The time of scientists is limited and we want to enable physicists, chemists, engineers and all others to focus on their domain. They want to optimizes the application regarding the scientific outcome, while our framework takes care of the deployment. We also want to encourage more scientists not to be afraid of such huge systems. The learning curve is high, if someone wants to use a Top500 supercomputer system. However, with EASEY, there exists a solution to use a more and more common praxis: Docker container. General purpose systems like the SuperMUC-NG are made for general purpose applications. With the presented performance in this section, we can substantially offer an additional deployment approach on those systems, for everybody.

 \section{Related Work}
 \label{sec:related}

The presented framework and its implementation bases on the development towards containerization and the abilities such encapsulated environments offer.

\paragraph{Charliecloud and Docker}

Priedhorsky and Randles from the Los Alamos National Laboratory introduced in 2017 in \cite{priedhorsky2017charliecloud} a lightweight open source implementations of a container framework: Charliecloud. The authors followed their basic assumptions that the need for user-defined software stacks (UDSS) increases. Dependencies of application's still need to be compiled on the actual target HPC systems since not all of them are available in the stack provided by the compute center. Todays and future users need particular dependencies and build requirements, and more over also portability and consistency to deploy applications on more than one system. This is offered by Charliecloud, which bases on Docker to build an UDSS image. 

The advantage of Charliecloud lays in the usage of the user namespace, supporting non-privileged launch of containerized applications. Within this unprivileged user namespace also all other  privileged namespaces are created without the requirement of root privileges. Therewith, any containerized application can be launched, without requiring privileged access to the host system, as described from Brayford et al. in \cite{brayford2019deploying}. In the same paper, the authors investigated the performance of Charliecloud scaling an AI framework up to 32 nodes. Their findings showed a similar, negligible overhead, although our performance analysis included more nodes. Concerning possible security issues the authors stated that Charliecloud is safe, since it  only runs inside the non-privileged user namespace.

Docker, described by Merkel in \cite{merkel2014docker} is considered an industry standard container to run an applications in an encapsulated environment. Nevertheless, since some containers require root privileges by default and others can not prevent privilege-escalation in all cases, as shown in \cite{bui2015analysis}, Docker is not considered a safe solution when deployed on shared host systems.

Besides Charliecloud and Docker a newer daemon less container engine attracts more and more attention. Podman, described in \cite{podman}, provides functionalities for developing, managing, and running Open Container Initiative containers and container images. Future work of this paper will include a substantial analysis of Podman and its possible enhancements for EASEY.

\paragraph{Shifter and Singularity}
Charliecloud was also compared to other approaches like Shifter and Singularity. The Shifter framework also supports Docker images (and others, e.g. vmware or squashfs), shown by Gerhardt et al. in \cite{gerhardt2017shifter}. In contrast, it is directly tied into the batch system and its scalability and security outside the initial cluster is not shown so far.

Also Singularity was developed to be encapsulated into a non-privileged namespace, security issues have been detected, for example in \cite{CVE-2019-11328}, where users could escalate the given privileges. An example is detailed in \cite{brayford2019deploying}. 

Choosing the right container technology is crucial, especially regarding the security of the host and other users. Since Charliecloud is considered secure and shows promising scaling behavior, we choose this technology for our framework, however, a Singularity extension might be added at a later stage.

Including Charliecloud in such a framework, only one related approach could be discovered so far: BEE.

\paragraph{BEE}

The authors of \textit{Build and Execution Environment BEE} in \cite{bee2018} propose an execution framework which can, besides others, also deploy Charliecloud container on a HPC infrastructure. Their approach focuses on a variety of different cloud and cluster environments managed by the same authority. This broad approach tries to unify the execution of the same container. Compared to EASEY, which aims to auto-tune the performance for Petaflop-systems, it does not include any optimization to the underlying infrastructure.

\textit{BEE} also includes a submission system for deployment of jobs, but deploys each single run command as one job. Our approach focuses on complex computations which might also include several steps within one \textit{job}. Regarding the data service, no possibility is provided in \textit{BEE} to connect to an external resource for data stage-in or -out. 

We consider \textit{BEE} as a valid approach to deploy the same container on many different technologies. However, we focus on auto tuning of adjusting screws to gain performance advantages, and our target infrastructures are high performance systems in the range of Petaflops and later Exaflops.

\section{Conclusion}
\label{sec:conclusion}

The presented architecture and its components are considered the starting point for further investigations. The heterogeneous landscape of computing and storage facilities need applicable and efficient solution for the next generation computing challenges. Exascale is only the next step as Petascale was a few years ago. We need to built assistent systems which can be adapted to the actual needs of different communities and enable those to run their applications on more and more different and complex hardware systems. On the other hand, building and using entire Exascale systems  will require enhancements in all pillars like fault tolerance, load-balancing and scalability of algorithms themselves. EASEY aims to enable scientists to deploy their applications today on very large systems with minimal interaction. With the ongoing research, we aim also to scale well on a full Exascale system in the future.

To close the gap between application developers and resources owners, a close collaboration is needed. In fact, the presented solution for an efficient usage of containerized applications with Charliecloud is only the first part. We need to go deep in the systems, optimize the hardware usage on a node or even CPU, accelerator and memory level and convince the scientific communities, that also their applications are able to scale up to such a level. Here, such an assistant system like EASEY, which automatically deploys optimized applications, will convince the communities and enable scientists to focus on their work.

\paragraph{Future Work}
\label{sec:future}

As mentioned throughout this paper, this version of EASEY is the first step towards a comprehensive framework to enable easy access to future Exascale systems. Those systems might have a hybrid setting with CPUs and accelerators side by side. EASEY will be extended to more layers which will also operate on the abstraction of the computing unit and introduce a code optimization which aims to optimize certain executions with more efficient ones adapted to the target system.

Porting originally CPU-based applications to such hybrid systems will require more research. Enable an efficient but easy to use approach will base on several layers, which functionalities and interfaces will be the main target of the future research question of this work.

The next direct steps focus on efficient data transfers, also including the model of data transfer nodes as investigated in the EU-funded project PROCESS and published in \cite{hluchy2020heterogeneous} and  \cite{processD51}. For such a data access also the authentication mechanism needs to be enhanced.

\subsubsection{Acknowledgment} 
The research leading to this paper has been supported by the PROCESS project, which has received funding from the European Union’s Horizon 2020 research and innovation programme under grant agreement No 777533.

%
%

\bibliographystyle{splncs04}
\bibliography{easey}

\end{document}